\documentstyle[12pt,psfig]{article}
\oddsidemargin--1mm
\begin{document}

\newcommand{\pd}{\partial}
\newcommand{\be}{\begin{equation}}
\newcommand{\ee}{\end{equation}}
\newcommand{\ba}{\begin{eqnarray}}
\newcommand{\ea}{\end{eqnarray}}
\newcommand{\mbf}[1]{\mbox{\boldmath$ #1$}}
\def\<{\langle}
\def\>{\rangle}

\begin{center}
{\Large\bf The wave packet propagation using wavelets}

\vskip 0.5cm
Andrei G. BORISOV ${}^a$ \ \ and \ \ Sergei V. SHABANOV ${}^b$

\vskip 0.5cm
${}^a${\it Laboratoire des Collisions Atomiques et Mol\'eculaires,\\
Unit\'e mixte de recherche CNRS-Universit\'e Paris-Sud UMR 8625,\\
B\^atiment 351, Universit\'e Paris-Sud, 91405 Orsay CEDEX, France \\
${}^b$\ Institute for Fundamental Theory, Departments of Physics\\
and Mathematics, University of Florida, Gainesville, FL 23611, USA}
\end{center}

\begin{abstract}
It is demonstrated that the wavelets can be used to considerably speed up
simulations of the wave packet propagation in multiscale systems. Extremely
high efficiency is obtained in the representation of both bound and
continuum states. The new method is compared with the fast Fourier
algorithm. Depending on ratios of typical scales of a quantum system in
question, the wavelet method appears to be faster by a few orders of
magnitude.
\end{abstract}

{\bf 1}. Owing to the fast development of the computational tools the direct
solution of the time-dependent Schroedinger equation has become one of the
basic approaches to study the evolution of quantum systems. Thus, the wave
packet propagation (WPP) method is successfully applied to time dependent
and time-independent problems in gas-surface interactions, molecular and
atomic physics, and quantum chemistry \cite{1,3,4,5}. One of the main issues
of the numerical approaches to the time-dependent Schroedinger equation is
the representation of the wave function $\left| \Psi (t)\right\rangle $ of
the system. Development of the pseudospectral global grid representation
approaches yielded a very efficient way to tackle this problem. The discrete
variable representation (DVR) \cite{6} and the Fourier grid Hamiltonian
method (FGH) \cite{7,8} have been widely used in time-dependent molecular
dynamics \cite{1,3,10}, S-matrix \cite{11,12} or eigenvalue \cite{13}
calculations. The FGH method based on the Fast Fourier Transform (FFT)
algorithm is especially popular. This is because for the mesh of $N$
points the evaluation of the kinetic energy operator requires only $NlogN$ 
operations and can be easily implemented numerically. In the standard FGH
method the wave function is represented on a grid of equidistant point in
coordinate and momentum space. It is well suited when the the local de
Broglie wavelength does not change much over the physical region spanned by
the wave function of the system \cite{13}. There are, however, lot of
examples where the system under consideration spans the physical regions
with very different properties. Consider, for example, the scattering of a
slow particle on a narrow and deep potential well. Despite the short wave
lengths occur only in the well, in the FGH method they would determine the
lattice mesh over {\it entire} \ physical space leading to high
computational cost. In fact, pseudospectral global grid representation
approaches are difficult to use in multiscale problems.

This is why much of work has been devoted recently to the development of the
mapping procedures in order to enhance sampling efficiency in the regions of
the rapid variation of the wave function \cite{10,13,14,15,17,18}. Though
mapping procedure, based on the variable change $x=f(\xi )$ is very
efficient in 1D, it is far from being universal. One obvious drawback is
that it is difficult to implement in higher dimensions. In the case of
several variables, there is no simple procedure to define the mapping
functions $f$ so that the lattice would be fine {\it only} in some
designated regions of space. Next, the topology of the new coordinate
surfaces can be different from that of Cartesian planes. Therefore the
Jacobian may vanish at some points (e.g., spherical coordinates). This leads
to singularity in the kinetic energy and imposes small time step in
simulations \cite{19}.

In this letter we propose a novel approach to the wave packet propagation in
multiscale systems. It is based on the use of wavelets as basis functions in
the projected Hilbert space. In contrast to the plane waves, the wavelets
are localized in both real and momentum spaces. This characteristic property
of wavelets allows us to accurately describe short wave length components of
the wave function in designated spatial regions, while keeping only few
basis elements in the bulk. This leads to a drastic reduction of the
projected Hilbert space dimension without any loss of accuracy. In some
cases it may even become possible to diagonalize the Hamiltonian matrix so
that the time evolution can be elementary followed in the basis of
eigenstates. We illustrate our approach by simulations of a simple
multiscale one-dimensional scattering problem. A generalization to higher
dimensions is straightforward and based on the standard mathematical
construction of multidimensional wavelets \cite{20}. Wavelet bases have been
successfully used for the systematic treatment of the multiple length scales
in the electronic structure of matter \cite{21a,21b,21c,21d}\ (for a
comprehensive review see \cite{21} and references therein). Here we apply,
for the first time, wavelets to study the time evolution in multiscale
quantum systems and demonstrate the advantages of the wavelet method over
the conventional FGH method.

{\bf 2}. To illustrate the efficiency of the wavelet method in the wave
packet propagation, we consider the 1D scattering of an electron on a narrow
and deep potential well. Despite we use an electron as a projectile,
procedure described below readily applies to molecular or atomic wave
packets. The potential has the form $\hat{V}=$ $-64\exp [-(100x)^{2}]$
(atomic units are used throughout the paper) and supports single bound state
with the energy $-0.63200$. We are interested in the transmission and
reflection coefficients for the energies of the scattered electron within $%
0.2-1.1$ range. This corresponds to a typical wave length $\lambda =2\pi $
which is much larger than the potential well width. To calculate the
scattering properties of the system we use the Gaussan wave packet impinging
on a potential well. The calculation requires a typical size of the box $%
x\in \lbrack -60,60]$, where 18 a.u. from each side are allocated for an
optical potential $\hat{V}_{opt}$ that absorbs reflected and transmitted
waves \cite{8}.

We first introduce a uniform lattice and simulate the wave packet
propagation by means of the FGH method. Split-operator technique \cite{7,19}
is used to calculate the action of the evolution operator $(\hat{W}=\hat{V}+%
\hat{V}_{opt})$:

\begin{eqnarray}
\left| \Psi (t+\Delta t)\right\rangle  &=&\exp (-i\Delta t\left[ \hat{H}+%
\hat{V}_{opt}\right] )\left| \Psi (t)\right\rangle   \nonumber \\
&=&\exp (-i\Delta t\hat{W}/2)\exp (-i\Delta t\hat{T})\exp (-i\Delta t\hat{W}%
/2)\left| \Psi (t)\right\rangle +O(\Delta t^{3})  \label{1}
\end{eqnarray}
Numerical convergence is obtained with $N=2^{14}$ points at the grid and a
time step $\Delta t=0.0002$. Calculated transmission and reflection
coefficients are shown in Fig. 1. The time evolution of the wave packet is
presented in Fig.2. The colors represent the magnitude of the wave packet in
the logarithmic scale: $-10\geq \ln |\Psi (t,x)|\geq 0.3$. As the color
changes from red to violet, $\ln |\Psi (t,x)|$ 
decreases from $0.3$ to $-10$. 
Without the potential present, we would see a colored ray (a trajectory of
a free particle is a straight line) spreading for larger values of $t$.

To compare the above results with our new wavelet method, we take the
Daubechies wavelets ${\cal D}_{10}$ which have ten first vanishing moments
and the filter of length 20 \cite{20}. The basis is generated by the scaling
function $\phi (x)$ whose support lies in the interval $x\in \lbrack 0,19]$.
The lowest resolution level is set so
that the initial wave packet is reproduced with high accuracy in the
orthonormal basis of the functions $\phi _{1,j}(x)=\sqrt{2}\phi (2x-j)$ 
where $j$ runs over integers inside the interval $(-120,120]$. 
The wavelets $\psi _{k,j}(x)=\sqrt{2^{k}}\psi (2^{k}x-j)$ are used
to describe short wave length components of the wave function in the
vicinity of the potential well. For a moment, they should be regarded 
as a special
orthonormal basis with compact support and the following properties. If the
basis of scaling functions $\phi _{1,j}(x)$ spans well functions whose
Fourier transforms are concentrated in a wave length band around $\lambda
_{1}$, the corresponding wavelets from the $k$th resolution level form a
good basis for a band centered at $2^{-k}\lambda _{1}$. In our case we have
used $k=1,2,3,4$. The number of wavelets needed on each resolution level is
determined by the potential width (by its shape, in general). The technical
details are explained in Section 3. It appears necessary to take 20 wavelets
on each resolution level. Thus, all together we use {\it only} $240+4\cdot
20=320$ coefficients (or, equally basis functions) to describe the wave
packet propagation instead of $2^{14}$ in the fast Fourier method. Due to
such a tremendous reduction of the size of the problem, the Hamiltonian
matrix can be directly diagonalized and the time dependent wave function can
be easily obtained as:

\begin{equation}
\left| \Psi (t+\Delta t)\right\rangle =e^{ -i\hat{V}_{opt}\Delta t}\,
 {\textstyle {\sum_{E}}}\, e^{-iE\Delta t}\left| E\right\rangle
\left\langle E|\Psi (t)\right\rangle \ ,  \label{2}
\end{equation}
where $\left| E\right\rangle $ and $E$ stands for the eigenvector and
eigenvalue of the Hamiltonian, respectively. Convergent results are obtained
with time step $dt=0.025$, i.e., much larger than in the fast Fourier
method above. The results are given on Figs. 1 and 3. There is a perfect
coincidence of the fast Fourier and our wavelet results, while the wavelet
method is nearly 100 times faster.

For the sake of comparison, we also used Eq. (1) to simulate the time
evolution. The same accuracy is achieved for a time step comparable to the
time step in FGH method. However there is an alternative, much better
splitting scheme thanks to the localization properties of wavelets.  
Consider the decomposition $\hat{H}=\hat{H}_{1}+\hat{H}_{2}$
where $\hat{H}_{1}$ contains matrix elements
of the basis elements localized in the vicinity of the potential well,
while  $\hat{H}_{2}$ corresponds to the free space and, therefore, 
contains matrix elements only of the scaling functions. 
The propagation can then be done accordingly: 
\begin{equation}
\left| \Psi (t+\Delta t)\right\rangle =e^{-i\hat{V}_{opt}\Delta t}
e^{-i\hat{H}_{2}\Delta t/2}\, \left\{ {\textstyle{\sum_{E_{1}}}}
e^{-iE_{1}\Delta t}\left|
E_{1}\right\rangle \left\langle E_{1}\right| \right\}
e^{-i\hat{H}_{2}\Delta t/2}\left| \Psi (t)\right\rangle\ ,  \label{3}
\end{equation}
where $\left| E_{1}\right\rangle $ and $E_{1}$ stands for the eigenvector
and eigenvalue of the $\hat{H}_{1}$. In the spit method (\ref{1}) 
the error depends
on the spectral range of operators involved. For larger spectral ranges,
the errors get larger. The conventional way to cope with this
problem is to reduce the time step. The approach (\ref{3}),
which we call the ``wavelet tower diagonalization'', offers a better 
alternative.
The convergence here is drastically improved because {\it (a)}
$[\hat{H}_1,\hat{H}_2]$ is small and {\it (b)} the operator $\hat{H}_1$
with a large spectral range can be diagonalized. Both the properties 
are hardly achievable without the wavelet basis.       
In our example the convergence is
reached for a time step $dt=0.01$ (vs $dt=0.0002$ in (\ref{1})). 
Corresponding results are
presented in Fig. 1. This approach applies even better
to evolving heavy-particle wave packets because the Chebyshev \cite{19,23}
or Lanczos \cite{19,24} schemes can be used. These schemes are, 
first, more efficient than the
split method and, second, allow one to take full advantage of the
sparse structure of the Hamiltonian in the wavelet basis (see below).

Note also that by making the potential depth greater so that the minimal
wave length becomes twice shorter, we would have to increase the lattice
size by factor two in the Fourier method, thus increasing the number of
coefficients needed to describe the wave packet by $2^{14}$ (!), while in
our wavelet approach one more resolution level is to be added, implying {\it %
only} 20 extra wavelet coefficients in the wave packet decomposition.
Thus, the wavelet approach  offers a systematic
and easy way to improve the accuracy of simulations. 

{\bf 3}. The description of technical details of our approach is limited to
essential practical steps necessary to reproduce our results and to apply
the algorithm to new systems. A general theoretical analysis of wavelets
bases in multidimensional spaces can be found in \cite{20}. So we only
discuss the Hilbert space $L_{2}$ of square integrable functions of a real
variable $x$, $\int dx|\Psi (x)|^{2}<\infty $.

{\it (i)}. The scaling function $\phi (x)$ is defined by the equation $\phi
(x)=2\sum_{l}h_{l}\phi (2x-l)$ (called the scaling relation), where the
coefficients $h_{l}$ are called a filter. For a finite filter, the
scaling function has compact support. Define $\phi _{n,j}(x)=2^{n/2}\phi
(2^{n}x-j)$ for all integers $j$ and $n$. The filter $h_{l}$ satisfies
an equation obtained 
by combining the scaling relation with the required orthonormality condition
of the scaling functions $\phi_{n,j}$ for fixed $n$. 
Given a filter, numerical values of $\phi (x)$
can be generated by an iteration procedure \cite{20,25}.

{\it (ii)}. The subspace of $L_{2}$ spanned by $\phi _{n,j}$ is denoted $%
V_{n}$. An important property $V_{n}$ is that $V_{n-1}\subset V_{n}$ and the
projection of any function $\Psi $ from $L_{2}$ onto $V_{n}$ converges to $%
\Psi $ in the $L_{2}$ sense as $n\rightarrow \infty $. Consider an orthogonal
decomposition: $V_{n+1}=V_{n}\oplus W_{n}$. There exists a function $\psi
\in W_{0}$ called the mother wavelet such that $\psi _{n,j}(x)=2^{n/2}\psi
(2^{n}x-j)$ form an orthonormal basis in $W_{n}$. The numerical values of $%
\psi $ can be generated from the scaling relation $\psi
(x)=2\sum_{l}(-1)^{l}h_{1-l}\phi (2x-l)$. A finite dimensional subspace of
$V_n$ is used to approximate $L_2$ in simulations. By construction,
the functions $\phi_{l,j}$, $l<n$,  
and $\psi_{k,j}$ for $k=l,l+1,...,n-1$ form
an orthonormal basis in $V_n$: $\left\langle \psi
_{k,j}|\psi _{k^\prime,i}\right\rangle =\delta _{kk^\prime}\delta _{ij}$,
$\left\langle \phi _{l,j}|\phi _{l,i}\right\rangle=\delta_{ij} $, and
$\left\langle \psi _{k,j}|\phi _{l,i}\right\rangle =0$.

{\it (iii)}. From the definition of $\phi _{l,j}$ and $\psi _{k,j}$ it is
clear that the index $j$ indicates the position of support of the scaling
function or wavelet. The index $k$ of $\psi _{k,j}$ can be understood as
follows. Let the Fourier transform of the mother wavelet $\psi $ be peaked
at a momentum $p_{0}$. Then the Fourier transform of $\psi _{k,j}$ is peaked
at the momentum $p_{k}=2^{k}p_{0}$. Since the wavelets with different $k$
are orthogonal, the coefficients $d_{k,j}=\left\langle \psi _{k,j}|\Psi
\right\rangle $ determine relative amplitudes of successively shorter wave
length components of $\Psi $ in the vicinity of $x=2^{-k}j$ as $k$
increases. For this reason, the index $j$ is called a position index, and $k$
is called a resolution level.

{\it (iv)}. From the physical properties of a system one can estimate a wave
length band, $\lambda \in \lbrack \lambda _{min},\lambda _{max}]$, required
for simulations. The lowest resolution level $l$ is identified by the
condition that $\phi _{l,j}$ span functions with Fourier components in the
vicinity of $\lambda _{max}$. The necessary maximal resolution level is
determined by $n\approx \log _{2}(\lambda _{max}/\lambda _{min})$. In our
example $l=1$ and $n=5$. If $N$ scaling functions are needed (to cover
a given physical volume), then, in
general, on every wavelet resolution level there will be $%
J_{k}=2^{k-l}N$ basis functions. The wave packet is decomposed in the
corresponding basis of $V_n$ 
\begin{equation}
\Psi (t)=\sum_{j=1}^{N}s_{1,j}(t)\phi
_{1,j}+\sum_{k=1}^{n-1}\sum_{j=1}^{J_{k}}d_{k,j}(t)\psi _{k,j}\ .  \label{4}
\end{equation}
The decomposition coefficients as functions of time are to be found from the
Schroedinger equation. Yet, the total number of coefficients in $\Psi (t)$
is about the same as that in the uniform finite lattice approach.

The advantage of using wavelet bases becomes significant whenever the volume
of regions where short wave lengths can appear during the time evolution is
much smaller than the total physical volume of the system. This allows one
to substantially reduce $J_{k}$ in Eq.(\ref{4}) 
by taking higher resolution level
wavelets only where they are needed. In our case we need only 20 wavelets at
each resolution level. We shall refer to all the wavelets needed in the
vicinity of one local minima of the potential as a {\it wavelet tower}.
Higher tower floors correspond to higher resolution levels $k$. Each wavelet
is thought of as a building block of width equal $2^{-k}$.

{\it (v)}. An important part of the algorithm is the projection of the
Hamiltonian onto the wavelet towers. Matrix elements of the potential as
well as the initial coefficients $s_{l,j}=\left\langle \phi _{l,j}|\Psi
\right\rangle $, $d_{k,j}=\left\langle \psi _{k,j}|\Psi \right\rangle $
are computed via Riemann sums. The matrix elements of the kinetic energy are
computed using the the fast Fourier transform to obtain the second
derivative of the function. The Hamiltonian matrix is sparse because of 
compact support of the basis functions. For instance, in our example 
$\left\langle \phi _{n,j}|H|\phi
_{n,i}\right\rangle =0$, if $|i-j|\geq 20$. Finally, the time evolution can
be computed by standard techniques such as split, Lanczos, or Chebychev
methods \cite{19,23,24}. If after the projection on wavelet towers, the
Hamiltonian matrix is small enough for a direct diagonalization, 
Eq. (\ref{2}) can
be used for the propagation.

In conclusion, we have demonstrated that  wavelet bases can
be extremely efficient for solving the time-dependent
Schroedinger equation for multiscale systems. The bound and continuum
states are accurately described with a much less number  of
basis functions as would be
required in the standard Fourier-grid methods. In our example we were able
to reduce the basis size by a factor of 50, which allowed us to scale down 
the computation time by a factor of 100. The method can easily be implemented in
higher dimensions where a wavelet basis is built by taking the direct
product of one-dimensional wavelets \cite{20}. The wavelet approach becomes
more efficient as ratios of typical scales get larger. Finally we would like
to stress that the wavelet method is ideally suitable for simulating wave
packet propagation in multiscale problems with complicated form of the
potential. This is because the wavelet towers can be custom designed for any
topology of the potential minima (e.g., potential valleys) 
regardless of the dimensionality of the system.

\vskip 0.5cm 
{\bf Acknowledgments}. S.V.S. thanks the LCAM of the University of Paris-Sud
for warm hospitality. We are also grateful to Victor Sidis and Jean Pierre
Gauyacq for support of this project, and to John Klauder for reading the
manuscript and useful comments.

\vskip 1cm
\begin{center}
Figure captions
\end{center}
Fig. 1. Calculated transmission (black) and reflection (red) coefficients.
Circles: The standard approach based on the split propagation and Fourier 
grid  with uniform mesh. Solid curves: The wavelet approach with Hamiltonian 
matrix diagonaliztion.
Triangles: The split propagation with the ``wavelet tower 
diagonalization'' as in  Eq. (\ref{3}).

\end{document}